# Internet TV: Business Models and Program Content

by

**David Waterman**
**Dept. of Telecommunications**
**Radio and TV Center**
**1229 E. 7th St.**
**Indiana University**
**812-855-6170**
**waterman@indiana.edu**

**Revised September, 2001**

**Paper for presentation to the 29th Annual TPRC Research Conference on Information, Communication, and Internet Policy, Washington, D.C., October 27-29, 2001**

This paper is forthcoming in Darcy Gerbarg (ed.), *IPTV* (tentative title), Lawrence Erlbaum Associates, 2002. An earlier version was published as D. Waterman (2001), The Economics of Internet Television, *Info: The Journal of Policy, Regulation and Strategy for Communications, Information, and the Media,* Vol. 3, No. 3. I am especially indebted to Andrew Odlyzko, Ben Compaine, Robert LaRose, and to participants at the Columbia Institute for Tele-Information "TV Over the Internet" conferences in New York (November 2000) and Dusseldorf (December 2000), for their comments.

# Internet TV: Business Models and Program Content

### Abstract

Internet technology should eventually provide important improvements over established media not only in the efficiency of broadband delivery, and of particular importance, in the efficiency of business models that can be used to collect money for that programming. I identify five economic characteristics of Internet technology that should lead to these greater efficiencies: (1) lower delivery costs and reduced capacity constraints, (2) more efficient interactivity, (3) more efficient advertising and sponsorship, (4) more efficient direct pricing and bundling, and (5) lower costs of copying and sharing.

The most successful Internet TV business models are likely to involve syndication to or from other media, and also international distribution. In the broader context, Internet TV is another syndication outlet by which program suppliers can segment their overall markets and thus support higher production investments. Many innovative and more sharply focused programs will surely prosper on Internet TV, but the attractiveness to audiences of high production value programming will tend to advantage broad appeal programming, such as Hollywood movies. Historical evidence about the performance of cable television and videocassettes is presented to support these points.

## I.  Introduction

Beginning in the late 1990s, the Internet unleashed an extraordinary amount of experimentation with the delivery of broadband entertainment content to consumers. Much of that content has been Internet-original, notably short films and serials, and interactive program forms. At the other end of the spectrum have been feature films and TV programs already appearing in theaters or on other media. The suppliers of this content have experimented with just as wide a range of business models: advertising, sponsorship, bundling with other products, promotion of other products, instant on-line purchase of merchandise, pay-per-view or "rental," outright sale of content (by consumer downloading), and hybrid forms. There has been a Wild West atmosphere to all these trials, with little consensus on what will eventually develop.

Amid much publicity, many of the cowboys have bitten the dust in the past one or two years.  But new experiments are still being launched, and few observers would discount the long term possibilities for delivering broadband entertainment over the Internet.  From that perspective, several big questions immediately come to mind.  What kinds of business models will predominate when Internet television eventually develops? How will file sharing technologies affect these business models? And what types of content will these business models support:? a wealth of new niches, or just more of the same?

Of course, no one can answer these questions with any certainty. Indeed, arriving at the answers is the market function of all the experimentation in the first place. My objective is much more modest. By applying some economic principles, and considering the historical experience with established broadband media, I hope to provide a useful framework for productive thinking about the answers to these questions.

### A.  Scope

To begin, some subject matter boundaries. First, I focus entirely on actual IP (Internet protocol) delivery of television programming. For example, "enhanced" television--using the computer (or a set top box) to interactively play along with a standard television exhibition of a sports event or *Who Wants to be a Millionaire?* does not count. Second, I concentrate mainly on dramatic entertainment forms for consumers. There are many business-to-business (B2B) broadband applications (such as videoconferencing), business-to-consumer electronic commerce that is unaffiliated with entertainment programming, interactive computer games or gambling, pornography, and news programming--all of which have evident economic potential on the Internet, but which are also outside my scope. Third, I am concerned with the long run future. Internet bandwidth capacity, as well as payment mechanisms and general usability of computers, will have to develop far more for Internet TV to reach its potential. I assume that those developments will eventually happen, but abstain on when. I also assume that in the long term the computer and the television set will converge. The basis of this assumption is my faith that if Internet TV entertainment technology develops in a potentially profitable way, the computer will find its way into living rooms. Finally, my own cultural limitations induce me to focus primarily on developments in the United States.

I begin in Section II below with a brief review of the academic literature relevant to Internet TV. Section III follows with an overview of recent Internet TV entertainment experiments. In Section IV, I set out five basic economic characteristics of the Internet in comparison with existing media that are relevant to development of Internet TV business models and content, followed in Section V with implications for future development. Section VI offers concluding remarks.

## II.    Literature

There is understandably little academic literature that addresses such new subjects as broadband entertainment content or business models on the Internet. Several books and articles, however, provide a foundation for the present study.

Owen (1999) takes a basically pessimistic view of the future of Internet television, arguing not only that adequate bandwidth appears to be far into the future, but that the architecture of the web is not well suited for broadcasting video. More recently, several papers in the communications literature have been concerned with Internet TV. Kiernan and Levy (1999) studied the content of broadcast related websites, for example, and a series of articles on the future of the Internet, notably Shaner (1998), have conceptualized the nature of web content on a broader level. Of more direct relevance to the present analysis, Picard (2000) offers a study of the historical development of business models for online content providers more generally, and suggests lessons in these experiences for the future. Konert (2000) analyzes a variety of financial and revenue generation models for Internet broadcasting from a European perspective, especially with regard to implications for European public broadcasting. Shapiro and Varian (1999) discuss economic characteristics of digital and networking technologies, including the Internet, from the standpoint of advising business people how to take advantage of these technologies and design better business models. Bakos and Brynjolfsson (1999) study the economics of product bundling strategies on data networks. A recent National Research Council report, *The Digital Dilemma* (2000) discusses in detail the economics and technology of the Internet from the standpoint of copyright and other government policy questions. A number of law review and other policy oriented papers, such as Samuelson (1999), Schlachter (1997), Einhorn (2000), and Jackson (2001) also discuss economic and technological characteristics of Internet entertainment delivery from those perspectives.

### III.     Internet TV Experimentation

A number of commercial websites offering Internet TV entertainment in the past few years illustrate the wide variety of business models and content under experimentation in the United States.

On the content side, film shorts and serials, or "webisodes," (also mostly very short) are very common fare on sites such as *Ifilmcom, atomfilms.com, shockwave.com*— and were on now-defunct sites such as *icebox.com, entertaindom.com* and *mediatrip.com*.[1]. Many of these programs have been originally produced for the Internet. The great majority of serials seem to be animation, though there are also talk shows (eg, *Cyberlove* on *thesync.com*). A few sites, such as *sightsound.com, Ifilm.com* and file sharing sites, such as *aimster.com* and *grokster.com*, offer access to full-length features. Most of these have been theatrical films, although one Internet-original feature distributed by *sightsound.com* in 2000, *The Quantum Project*, received a lot of publicity. *Signtsound* has now begun offering recent full length features from the Miramax studio. In the past two months, two consortia of Hollywood studios have announced that they will very soon offer their major films over the Internet. One now defunct site, *itsyourmovie.com,* offered an original serial program in which viewers could vote on the future plot direction. Sony's site, *screenblast.com,* allows users to interactively create their own mini-episodes of *Dawson's Creek*. Many broadcast TV stations now stream streaming their programming from websites, and *broadcast.com* retransmits television stations worldwide. Many other sites, such as *CBS.com* and *NBCi.com*, have offered a wide variety of short video news clips, previous episodes of entertainment series programs, and of particular interest, outtakes or other original clips that supplement regular TV series (eg, CBS's *Survivor*). Movie and TV show previews or other promotional videos for entertainment programs are very common fare on entertainment websites.

---

[1] *atomfilms.com* and *shockwave.com* merged in early 2001 (*AtomShockwave*), and have subsequently announced major layoffs. *Ifilm.com* has recently been reported to be on weak footing,

The variety of business models employed by these Internet broadband sites has been great. Banner advertising and its increasingly pro-active forms, other links to retail outlets, and on-site merchandising have become common components. Brief "pre-roll" commercials now appear before some webisodes. Users can click on these commercials and get more product information. Some sites, such as *BMWfilms.com* and *skyy.com,* now offer high production quality short films that overtly promote products their sponsors' products (namely BMWs and Skyy Vodka) by integrating them into the stories. Instant on-line purchase of products modeled within an entertainment program was apparently first experimented with by Microsoft. Internet TV promos of programming available on other media, especially broadcast or cable programs and movies, have become common on the Internet. Among the most substantial websites that has pioneered a direct payment model is *sightsound.com*. This site offers buyers the opportunity to either "rent" (obtain a one or 2 day license) for about $3.95, or "buy" (ie, download) recent and older theatrical features, for about $14.95. *Sightsound.com* also sells videos and DVD's by mail, and has bundled exhibitions of *Quantum Project* with a free download of media player program. As of Fall, 2001, several Hollywood studios have announced consortia that will offer their movies on a pay-per-view or "rental" basis over the Internet.

IV.     **Five Economic Characteristics of the Internet affecting Internet TV Business Models and Content.**

The idea that the Internet is a revolutionary communications medium has become common currency in discourse about the media. This label is certainly justified in some contexts. From my perspective, though, the Internet is best viewed in comparison with established broadband media in terms of the economic *improvements* it can make to cost and efficiency features of those media.

I divide the Internet's economic improvements upon established media into five categories: (1) lower delivery costs and reduced capacity constraints, (2) more efficient interactivity, (3) more efficient advertising and sponsorship, (4) more efficient direct pricing and bundling, and (5) lower costs of copying and sharing.

### (1) Lower delivery costs and reduced capacity constraints

Media transmission system costs consist of several components: a capital infrastructure for transmission, home premises equipment, and variable costs of delivering the information. Parts of these infrastructures and home equipment have multiple uses, and costs often depend critically on usage rates. Cost comparisons among media are thus difficult. Some comparisons show that Internet transmission of television signals is currently far more expensive than cable and some other media (see Noll, 1997; Noam, 2000). As Internet broadband capacity develops, Internet TV transmission will certainly become cheaper. Internet TV is also more-or-less free of geographic constraints, allowing essentially instantaneous worldwide transmission. A component of delivery costs is the ability of consumers to simply download content rather than to copy in real time off of a cable channel, for example, or take a trip to the video store to buy a product that has been manufactured, packaged, shipped, and maintained in an inventory. From the latter perspectives at least, Internet-transmission of video is quickly becoming more cost efficient than existing media.

Because of its architecture, capacity or "carriage" constraints become very minor on the Internet. In the 1940s, increasing the number of available movies in a town meant building a whole new theater. Broadcast TV stations reduced these capacity costs, especially in larger markets. Cable TV and DBS have further reduced capacity constraints, and these costs continue to fall with digital compression technologies. Video stores have essentially the highest "capacity" of any established broadband media. All of these media, however, have significant carriage costs. Another channel on a cable system requires a major investment, even with digital compression technologies. Another video or DVD at retail stores requires total demand for a few thousand copies to make duplication and physical distribution worthwhile. The stores that carry each title must cover inventory costs for as long as consumers wish to rent or buy it. On the Internet, a variety of websites can offer a virtually unlimited number of products, and consumers can readily switch within and among different sites.

The implication of these cost and capacity advances is lower prices and especially, greater product variety. That variety provides one ingredient for virtually "true" video-on-demand systems. Also, thinner and more marginal markets can now be served.

The latter potential is shown by the abundance of Internet-original short films and serials already available. One factor is probably just their suitability to a medium in which more lengthy viewing or downloading experiences are now too tedious. Many of them have earned critical praise, though, and their often racier content is generally differentiated from other broadband content. Nevertheless, another economic reality underlies their prevalence on Internet TV; consumer demand for short subjects has in the past usually been too marginal for all but a few to even be made available in specialty video stores or on the most narrow appeal cable television channels.

Another example of relatively marginal content on the web is movie or program outtakes: footage about the making of a program, interviews with the creative people, etc. Currently, such material is included as extras on some DVDs, but as some websites are already demonstrating, the Internet expands these possibilities almost without limit. Similarly for supplementary material about advertised products. Such ancillary video materials are important building blocks for both advertising and direct pricing business models.

### (2) More efficient interactivity

If the Internet has a forte among its many marvels, it is surely 2-way interactivity. Interactivity has been physically possible since cable systems offered it in early years, notably on the QUBE system in Columbus, Ohio in the 1970s. Also, a hybrid form of interactivity is now available with the integration of computers and standard TV transmissions to create enhanced TV. Viewers with a computer in the same room (or a set top box) can simultaneously play along with game shows or sports events. Viewers can also buy products shown on standard television commercials more and more easily with the right home equipment. In some experimental systems, viewers can now choose between several simultaneous feeds of standard broadcast content (such as different camera angles covering a sports event) to control the pictures that they actually view. PVRs (personal video recorders) permit asynchronous control of programming starts and stops.

Cable, DBS and other multi-channel systems are rapidly developing interactive technology as well. But the development of Internet TV should permit most of these activities to be conducted more efficiently. Viewers can instantly and easily control a much wider variety of programming content via their responses. Home shoppers can simply click on a product shown in the middle of a televised movie to instantly buy it, or to get more information about it. They can do the same with an in-show TV commercial. More efficient interactivity offers another ingredient of true video-on-demand systems as well: a convenient process of ordering movies or other programs for on-screen viewing or for download.

With Internet technology, viewers can also neatly manipulate the sequencing of video images they see. Many question whether consumers in any significant numbers will (apart from the case of pornography) ever want to fiddle with the narrative form of entertainment programs. Still, a great amount of innovation is being invested in systems that will allow them to do that if they want.

**(3) More efficient advertising and sponsorship**

A perennial limitation of television advertising has always been waste circulation because of muddy demographic, product interest, or other segmentation. Cable and other multi-channel systems have reduced this problem by making room for more and more sharply targeted programs. Internet television permits the chance to further advance this quest in two ways. First, the virtual removal of capacity constraints should allow still sharper segmentation in the same way that multi-channel systems have improved the broadcast model. Second, the ability of advertisers to track the buying or Internet usage patterns of individual consumers permits different ads to be inserted within (or displayed alongside) the same program, depending on the viewer's revealed interests or estimated willingness to buy a particular product.

The click-through interactive system of Internet advertising is much like the "per-inquiry" (PI) ads often seen on emerging cable TV networks, in which the network is paid not for exposures, but earns a % cut of each purchase made via a phone number displayed on screen. The Internet system is simply a more efficient PI system. Even without click-through purchasing opportunities, the ease of obtaining more information about products with a mouse-click is a significant advance in production information dissemination. Finally, the Internet offers the opportunity for full sponsorship of a website, or of an area within a site, that attracts consumers with entertainment programming. This sponsoring system might meld the branding of a dramatic format program and its characters with a consumer product in a better way than television program sponsorship, first developed in the late 1940s, has been able to do in the past.

Although Internet technology thus promises more efficient advertising based-business models to support broadband programming, history suggests formidable practical limits. First, although multi-channel cable television had brought forth more than $13 billion in total advertising by 2000,[2] including many new advertisers, the "magazine model" of higher rates for sharper segmentation has not materialized. A few networks, such as MTV, have segmented very successfully, but cable network cost-per-thousand ad rates are on average still about 25% below those of the major broadcast networks, apparently due mostly to the limited national audience reach of networks that rely on multi-channel system delivery. (Media Dynamics, 1999; Waterman and Yan, 1999). Second, while in-show commercials can be carried on Internet TV programs, the click-through potential on the Internet does not seem to offer a great advantage over what virtually ubiquitous broadcast television stations already do with in-show commercials, especially given the relative importance of product image advertising. Most products are not subject to impulse purchase or PI models of advertising. Third, as we have seen with VCRs and are beginning to see with PVRs, consumer control digital technologies like the Internet generally increase the ease with which viewers can zap ads, or otherwise avoid them.

Overall, the success of advertising as support for Internet broadband entertainment seems to rely heavily upon consumer initiative to investigate or make on-line purchases of advertised products--a plausible model, but one with a spotty historical record on other media. Furthermore, to the extent Internet TV evolves into a "store and replay" rather than "live" transmission medium, as some believe will happen, advertising's potential will also be limited (see Odlyzko, 2000).

---

[2] National Cable Television Association (*www.ncta.com*)

In the past year, analysts' expectations for the potential of Internet advertising models have greatly diminished--for the above and perhaps other reasons. Nevertheless, innovation is active, and at least some of the potential improvements to advertising efficiency on the Internet should materialize. For at least some products, the result should be more cost-effective advertising and product promotion, and thus an increase in the effectiveness with which advertising and sponsorship can support broadband entertainment content. More sharply focused programs should accompany these developments.

### (4) More efficient direct pricing and product bundling

More efficient direct pricing means lower costs in making transactions, but especially the ability to more effectively price discriminate—that is, to extract the maximum amount that each consumer is willing to pay for a product. In several respects, Internet technology promotes these efficiencies.

First, direct payment-supported video-on-demand systems are likely to be at least as cheap and easy to manage by web sites as they ever will be on cable or DBS. Internet payment systems now in development will presumably make payment as simple and secure as adding charges to a monthly cable bill. Micropayments, which allow very small amounts (perhaps only a few cents) to be automatically charged to a user via a credit card or similar means, are a prospective component of true video-on-demand systems. Micropayment systems are now in use for some Internet applications and are reported to be in development for a number of others.

Web sites can also engage in so-called dynamic pricing, by which consumers are charged different prices according to their perceived willingness to pay, based on prior purchasing habits on the web, website visiting habits, or other information. Basically, dynamic pricing permits more efficient price discrimination through better identification of high vs. low value customers.

An important component of effective direct payment systems is efficient bundling of products, such as a package of three movies together, monthly subscriptions, or the sale of movies along with talent interviews, outtakes, etc. A large literature in economics has explored many ways that such packaging can extract consumer surplus via price discrimination. (eg, Adams and Yellin, 1976; Varian, 1989).. Of course, video stores and cable or satellite based systems also offer bundles. But on the Internet, tailor-made packages can be offered to different consumers depending on buyer profile data, and interactivity allows choice among more complex menus or package variations than other media can efficiently offer.[3]

A variety of other price discrimination devices, such as reduction of prices over time for movies as they become older, or lower prices for repeat viewings, are also efficiently managed on the Internet. A plausible method of Internet TV price discrimination may involve consumer segmentation based on demands for different qualities of transmission. Consumers with higher speed connections, for example, are likely to have higher valuations for high technical quality.

As with advertising-based business models, these potential improvements in direct-pricing have practical limits. There is a long history of apparent consumer resistance to paying at every turn (e.g., the failure of DiVX, the digital video disk system promoted by Circuit City in the U.S. that allowed consumers to pay according to the number of times a program was watched). More generally, pay-per-view systems have not done very well on cable or satellite systems, although it is unclear how much the lack of consumer control over starts and stops, the limited selections, or other factors are responsible. Dynamic pricing may also have an uncertain legal future. Also, even though the Internet theoretically allows practically any kind of segmentation to take place, it may also prove difficult to price discriminate geographically with an inherently nationally and internationally distributed medium. Geographic discrimination is a natural process for video stores and cable systems.

Undoubtedly, some potential aspects of direct pricing on Internet TV will never happen. The Internet offers such potential in this area, though, that at least some of its advantages, in terms of lower transactions costs and more efficient market segmentation, seem bound to materialize. The result should be more viable VOD systems and greater revenue support for products with relatively high consumer demands.

---

[3] Bakos and Brynjolfsson (1999) study the economics of offering menus of very large bundles on the Internet.

### (5) Lower costs of copying and sharing

Attracting more recent attention than any other attribute of the Internet is the remarkable ease with which content—including movies or other videos—can be duplicated and transferred from one consumer to another. Popularity of *napster.com* and *gnutella.com* file sharing systems are a testament to these efficiencies. The limited use with video content on these systems thus far is no doubt due largely to bandwidth constraints. Of course, copying and sharing of movies and other videos has been widely practiced since VCRs arrived along with copy-prone pay-TV movies and prerecorded cassettes that can be copied back-to-back. The consumer's task of copying and sharing simply becomes far less time consuming and awkward with the use of a computer.

As everyone has recognized, computer network technologies for copying and sharing pose a serious threat to copyright holders because paying customers can practically evaporate from the market. Even a single casual file transfer can have devastating cumulative effects as it is retransmitted from user to user virtually without cost.

Attracting increasing attention are the new opportunities for copyright owners created by efficient duplication and file sharing via the Internet. I have already mentioned the negligible costs of a consumer download—essentially equivalent to copying--compared to purchasing a DVD or videocassette, or of making a real time copy off of standard television. With existing pay-per-view or home video systems, consumers who want to share a copied movie with someone else also have to physically deliver it to the recipient. Peer-to-peer computer transfer essentially eliminates that cost.

Fundamentally, the lower consumer costs of copying and peer-to-peer transfer via the Internet create market value. If distributors can manage to appropriate some or all of that created market value, their revenues and profits will rise (Besen, 1986). Consider the "old" system in which consumers have made real time back-to-back copies off of prerecorded videos or off pay-per-view channels to share with others. The copyright owner may be able to appropriate some fraction of the value of that physically shared copy to the recipient, but it is almost certainly lost revenue for the most part.[4] If we assume for the moment that distributors are able to maintain strong copyright protection governing broadband Internet transmissions, they may be able to appropriate all, or at least a larger part, of the value of an electronically shared copy. For example, an automatic electronic payment to the distributor could be activated by a peer-to-peer file transfer (such as via a *gnutella* or *napste*r-like system) of a copyrighted movie.[5] Alternatively, such peer-to-peer file transfers could be forbidden by copyright owners, and all users simply induced to purchase directly from the owner. In these eventualities, the incentive for consumers to engage in the cumbersome process of physical copying and sharing will also be reduced to the extent that prices for authorized electronic download or peer-to-peer file sharing are low enough to render the physical process a less desirable alternative.

---

[4] Besen makes the unrealistic assumption in his model that the distributor can appropriate all of this value. In fact, the most that the distributor can ordinarily appropriate is the value of the product to the buyer plus

Internet technology thus increases the distributors' potential revenues from movie or other product distribution. Possibly, these revenues can be enhanced by improved price discrimination as well. Those who take advantage of file sharing probably tend to have lower price demands, and thus may drop out of the market at the distributor's price for the "original" movie. If distributors can devise a method for charging lower prices for movies transferred from peer-to-peer file sharing sites, or from other sites that involve greater consumer search costs, than for direct downloads from the distributors' sites, they can also increase revenues.[6]

---

the value that buyer realizes from making and distributing copies. The latter component is likely to be less than the value of the copies to those who receive them. See also Katz (1989) for a useful analysis of home copying issues from an economic perspective.

[5] One indication of this potential is that the movie site, *sightsound.com* has reportedly been using *gnutella.com* to deliver encrypted movie files to users, who in turn pay *sightsound* a fee for the key (Snell, 2001).

[6] Essentially, the recent negotiations between *napster.com* and Bertlesmann for *napster* to price their music services to consumers were headed for just such a price discrimination system. According to press reports, a $2.95 to $4.95 monthly subscription price was reportedly being discussed for a fixed number of music file transfers on *napster*. For $5.95 to $9.95, unlimited transfers could be made. Additional charges would be made for the right to record the music onto blank CDs. The technical quality of all these paid subscriber transfers or recordings, however, would only be "near-CD" quality. Thus, higher value consumers would be induced to pay progressively more to use the service, but restrictions on transmission quality would still serve to segment the higher value CD and lower value file-transfer market segments. (Clark, 2001). Of course, the well-publicized resistance of other record companies to Bertlesmann's proposals suggests that such a direct pricing system was too clumsy or impractical in the current environment. As of Fall, 2001m napster was reported to be still planning to convert to a pay service, perhaps with a number of the above elements intact.

Of course, it is unrealistic to believe that e-mail or other unpaid peer-to-peer transfer of movies and other broadband entertainment could ever be eliminated, even if these practices were made illegal. Also, Internet distribution of movie data stripped from DVDs remains a serious threat. However, watermarking and other copyright protection technologies for authorized Internet distribution are rapidly developing, and the recent entry of Hollywood studios into Internet distribution of their movies suggests improved technologies of protection. If copyright interests continue to get favorable court interpretations of the 1999 Digital Millennium Copyright Act's prohibition on attempts to defeat encryption, and new legislation is enacted to account for newly developing problems, copyright owners should be able to keep losses to a minimum.[7].

The historical experience with back-to-back video copying by consumers encourages that speculation. Surveys indicate that consumer sharing of back-to-back video copies accounts for only about 1% of legitimate market transactions, and that the overwhelming proportion of consumers believe that back-to-back copying is illegal, suggesting that relatively small minorities would try to defeat encrypted programming even if they could, or would make illegal peer-to-peer transfers.[8] Those who do are likely to be low value consumers who would be disinclined to pay for the programs at the prevailing retail prices in any case.

In summary, Internet technology offers many ways by which program distributors can not only reduce costs of delivery and improve desirability of the programming packages they offer to consumers, but also improve the advertising, direct pricing, and other components of their business models. Web entrepreneurs are already combining components of business models in imaginative ways. Undoubtedly, some of these potential improvements will not work out. The law, the advance of technology, and uncertain demand could all inhibit them. But the potential of the Internet seems so great in these respects that it is hard to imagine that Internet TV will not—at least eventually-- lead to some dramatic improvements in television distribution and the business models that support it.

---

[7] See Jackson (2000) for a concise discussion of the DMCA and its application to the na*pster* case.
[8] The video copying percentage is derived from Office of Technology Assessment (1989) and Macrovision, Inc. (1996). The Macrovision study also reported that over 95% of survey respondents said they believed back-to-back copying of prerecorded videocassettes is illegal.

## V.    Implications for business model and content development.

### A.    Advertiser vs. direct pricing support

Internet television is disproportionately reliant on advertising or e-commerce related business models now--although no one seems to claim profits to date with any model. A shift toward direct-payment models already seems underway, and as bandwidth capacity improves, that trend is likely to continue for two reasons. One is that the intense competition among web distributors to establish themselves in the market during the Internet's growth stage has surely inhibited many firms from charging directly. The second reason to expect more direct pricing is that higher bandwidth capacity will mean that products of greater consumer value, namely feature length movies and sporting events, can be attractively presented. Other than pornography, consumers have never been willing to pay directly for much audio/visual entertainment besides movies and some sports. Historically, advertising has mostly been used to support content watched by low value viewers who are unwilling to pay enough to outdo the few cents per viewer that advertisers will pay for an exposure. Although more efficient advertising and e-commerce related systems are likely to increase the value of Internet exposures to advertisers, it seems unlikely that these improvements will overcome the basic economic forces guiding high value viewers toward direct payment systems.

### B.    The Internet as a component of multi-media and international syndication models.

As broadband media have proliferated in the past two or three decades, individual programs are more and more frequently distributed on several different media over a period of time. As countries throughout the world have privatized their media and relaxed trade barriers since the mid-1980s, international markets, especially for U.S. entertainment products, have also expanded.

As Internet TV develops, there will be tremendous economic pressures for the providers of its content to employ similar multi-media syndication models, as well as to supply products that have international appeal. In brief, higher revenues can be generated both because total potential audiences can be reached and because audiences can be more efficiently segmented. The result is greater potential revenues, which will support higher production investments, and in turn attract larger audiences. Another factor favoring multi-media syndication is marketing. A high expenditure ad campaign supporting a product release on one medium serves to increase demand in all the products' potential syndication markets, and thus realize economies of scale in the same way that high production investments can be spread over large potential audiences.

The best illustration of the compelling economic logic of multi-media syndication models is the current system of theatrical feature film distribution.

### 1. Movie distribution and Internet TV

We are all generally familiar with the process in which movies are released over time in sequence to theaters, then to hotels and airlines, to videocassettes and DVD, to pay-per-view television, to monthly subscription pay TV, and finally to television broadcasting or basic cable networks. It is widely recognized that this release sequence is basically a method of price discrimination.

The key requirement for any price discrimination is the ability to segment high value from low value consumers. The movie release sequence appears to involve two main segmentation devices. The first is time separation between release to different media. High value consumers having intense demand for a particular movie (or movies in general) are induced to pay higher prices for a first run theatrical exhibition, while other viewers wait for video, pay TV, or later exhibitions. The second segmentation device in movie distribution is product quality. In general, a theater offers a higher quality exhibition than does a TV exhibition. Similarly, the ability of a VCR or DVD player to stop and start a movie, the absence of commercials on PPV, etc. are quality attributes that attract higher value consumers. The end result is that effective prices paid by different consumers in the release sequence tend to drop over time.

While it has been remarked (sometimes gleefully) that the business models of movie distributors will have to change and adapt with the Internet, Internet TV actually fits naturally into this ready-made model. If effective unbundled direct pricing models evolve, and piracy of the Internet distributions themselves does not prove overwhelming, Internet movies will probably be exhibited in a similar window to that currently occupied by PPV or by video rentals and sales. Internet advertising models will probably be less valuable to movie distributors, but to the extent they do prove efficient, movies can be released on the Internet with advertiser or other commercial support, presumably at a later point in their business life. Precisely where the Internet fits into the movie distributors' business models depends on uncertain technological, legal, and demand developments, and will evolve from experimentation.

Wherever the Internet eventually fits, it is unlikely to replace other movie media in the sequence. All of them, from video stores to pay cable systems, have different quality attributes or different demographic appeals that further the distributors' objectives of segmenting markets in order to charge different prices for essentially the same product. DVD or videocassette retailers, for example, offer services that may never be effectively duplicated by the web. The physical search and human interaction in shopping for videos may have inherent advantages, as do the joys of physically owning a professionally packaged DVD or a tape. A related advantage of retailers is gift marketing, for which a well-packaged physical object is highly valued.

Theatrical film distribution also demonstrates the advantage of multi-media marketing. Advertising and publicity campaigns sometimes costing as much or more than the distributor grosses from theatrical exhibition are launched to support a theatrical release. Much of the benefits from this campaign are reaped as the film travels to video, pay TV, and other media in the subsequent months and years.

## 2. Syndication of Internet-original and other products

The market segmentation/price discrimination opportunities for multi-media syndication are not confined to movies. Many programs, including broadcast network programs, made-for-pay (monthly subscription) movies and series, direct-to-video features, and made-for-(basic) cable programs, all depend heavily on syndication to other media, in domestic and foreign markets, and they maximize their revenues by similar means of market segmentation.

Along similar lines, there is certain to be a wealth of "Internet-original" programming that is exhibited later, or even simultaneously, on other media. The business model of *AtomShockwav*e, for example, is already heavily dependent on multi-media syndication. In addition to its website-based advertising, *atomfilms.com* distributes collections of its best short films to cable networks, airlines, and other media, and compiles them onto DVDs and videocassettes for rental and sale as well (Long, 2000). It was estimated in late 2000 that atomfilms earned two-thirds of its total revenues from these "off-line" sources (Mathews, 2000).

### C.      Empirical Comparisons

#### 1.  Multi-market syndication and programming budgets.

The resulting economic advantage of program syndication over time is simply that larger program budgets, and thus programming with higher production values, can be supported. These more expensive programs attract larger and higher paying audiences. The contrasts in program investments of various entertainment products in the U.S. are illustrative. While there can be wide variance, the average major Hollywood studio theatrical feature was reported to cost about $55 million in 2000 (MPA, 2001). Based on recent trade reports, HBO's made-for-pay feature films average something over $8 to $10 million, made-for cable and made for broadcast features cost $3.5 to $5 million, one hour network dramatic series average $1.5 to $2 million per episode, a basic cable drama averages $750,000 to $1.2 million per episode.[9] All of these programs depend heavily on aftermarket syndication to support their investments levels, and there is a general correspondence between these programs' budget levels and their viability in aftermarkets.

By contrast, a recent report in *Variety* states that 3 to 5 minutes webisodes are budgeted at approximately $10,000 to $20,000.(Graser, 2000). Some Internet-original long feature projects have been reported to cost in the $80,000 to $100,000 range. ***Quantum Project*** is evidently the most expensive made-for-Internet film to date at approximately $3 million for about 36 minutes of entertainment, and its goal was clearly to gain publicity for *sightsound.com* (Cheltwyn, 2000). Of course, these relatively low budgets partly reflect the currently low household penetration rates of broadband Internet capability. They emphasize the point, however, that while creativity and imagination can go a long way on a shoestring, and sometimes lead to extraordinary results, the most successful Internet television programs are likely to be those that can be successfully adapted and sold to other media.

---

[9] These data compiled from *Variety*, March 6, 2000-March 12, 2000, p. 58: TNT taps DeBitetto Originals Prexy; Schneider, et al (1999), The Green Behind the Screen, *Electronic Media*, August 2.

## 2. The videocassette and cable experience

The economic significance of syndication-based business models is illustrated by the experiences of prerecorded videocassettes/DVD and cable television in the U.S.

Video content, as described by the time usage patterns for home video in Section (a) of Table 1, is dominated by feature films and children's programs. As a visit to any video store shows, a vast number of obscure, narrow appeal movies, how-to, and other programs are also available on video, but as Panels (b) and (c) of Table 1 suggest, the overwhelming portion of revenues are generated by the theatrical feature films of major distributors. A small proportion of the feature films on video are direct-to-video movies, but these are often syndicated to cable or broadcast television. Most of the children's programs on video are also exhibited on cable or broadcast television . A very small proportion of video content relies solely upon video rentals and sales for revenues.

Available data for cable television in the U.S. is reported in Table 2 is badly out of date, but also suggests the economic importance of multi-market syndication. Theatrical features dominate pay cable networks, and among the minority of originally-produced programming on pay cable networks, including made-for-pay feature films, a large percentage of that is no doubt later released on video, on broadcast or basic cable channels. For basic cable networks, the proportions of originally produced programming are much greater. Aftermarket feature films and off-network programs accounted for less than half of viewing in this study, but for dramatic programming formats, the overwhelming portion of viewing was directed to aftermarket programming. The percentages of original dramatic and other entertainment programming on basic cable may have increased since the mid-1980s, but again, a substantial percentage of that programming later ends up on broadcast channels.

Because the same basic economic forces are at work, there seems good reason to expect multi-market syndication business models also to dominate Internet television.

### VI. Conclusion: New niches vs. broad appeal programs on Internet TV

The widely discussed opportunities for interactive and other new and innovative Internet entertainment, and for more narrowly focused and marginal programming in general, are backed by economic logic. Lower costs, virtually unlimited capacity, efficient interactivity, and more efficient business models will all contribute to making them possible. As the video and cable TV experiences suggest, however, there will also be powerful economic forces favoring relatively expensive, broad appeal programming, such as Hollywood movies, on Internet TV.  By their nature, those are the types of programs that are most amenable to syndication on a variety of different media, as well as amenable to lucrative worldwide distribution.

A major challenge to the suppliers of Internet-original television programs will be to find lucrative aftermarkets for them, both in their domestic and foreign markets around the world. Inherently, these economic pressures also tend to encourage homogenization of content, as well as to limit the budgets of programs that have few alternative outlets. For example, interactive programming that depends on Internet architecture, or raunchier productions that do not adapt well to other media, will have major budget handicaps to overcome. It is a good guess that like basic cable TV, Internet TV programming will evolve into a dichotomous mix of niche-oriented, but relatively cheap Internet-original fare on the one side, and mass appeal, relatively expensive multi-market syndicated programming, on the other.

Few observers would claim that the diversity of entertainment programming, including much new and original, more sharply focused content, has not been greatly enriched by home video and cable television. But the results do seem to have fallen short of many of the hopes that visionaries' had for these media—spectacularly so in some cases, such as the grand hopes for "high culture" performing arts.

Historical experience suggests to me that these outcomes fell short of aspirations for three reasons. First, other things being equal, focusing program content on particular interests of small subsets of people seems to have stimulated demand (of both audiences and advertisers) less than was imagined. Second, visionaries underestimated the audience-drawing power of high production values. By spending more and more on the best stars, locations, and special effects, and spreading those costs over a potentially very large, multi-media audience, producers have been able to keep the lion's share of the viewers. Finally, many have underestimated the power of effective marketing. To support an opera in the United States certainly becomes more feasible with greater capacity (even if it can attract no more than 1% of the country's television homes), but an opera cannot compete with a blockbuster movie or a boxing match that can realize economies of scale in a national marketing campaign.

For some combination of all these reasons, its seems, the opera and other niche audiences have so far decided in the end to watch *Titanic*.  So far, Internet-original entertainment programming shows great creative promise, but the economic challenges will not be easy to overcome.

.

**Table 1**
**Videocassette/DVD content data**

| Time of use By content (1997) | | Boxoffice Market Shares (1998) | | Video Shipments Market Shares (1998) | |
|---|---|---|---|---|---|
| Feature films | 81% | 7 major studios | 87% | 7 majors studios | 83% |
| Sitcoms | 1 | Independents | 13 | independents | 17 |
| Drama series | 1 | | | | |
| Childrens | 12 | Total | 100% | Total | 100% |
| Sports | 3 | | | | |
| Other | 2 | | | | |
| | | | | | |
| Total | 100% | | | | |

Sources: (a) Media Dynamics, (b), (c): Paul Kagan Associates

**Table 2**
**Cable Television Program Content by Source**
**1986**

| | All programming | Dramatic only |
|---|---|---|
| **Premium channels** | | |
| | | |
| Original | 15% | 8% |
| Off-network | 2 | 2 |
| Theatrical film | 83 | 90 |
| Foreign acquisition | - | - |
| | | |
| Total | 100% | 100% |
| | | |
| **Basic channels** | | |
| | | |
| Original: | 56% | 3% |
| Off-network | 30 | 63 |
| Theatrical film: | 12 | 32 |
| Foreign acquisition | 2 | 2 |
| | | |
| Total | 100% | 100% |

Source: Waterman and Grant, 1991